\documentclass
[preprint,showpacs,byrevtex,10pt,twocolumn,tightenlines,prl,letterpaper]{revtex4}%
\usepackage{amsfonts}
\usepackage{amsmath}
\usepackage{amssymb}
\usepackage[dvips]{graphicx}
\usepackage{hyphenat}%
\providecommand{\U}[1]{\protect\rule{.1in}{.1in}}

\begin{document}
\preprint{ }
\title{Transition to exponential relaxation in weakly-disordered electron-glasses}
\author{Z. Ovadyahu}
\affiliation{Racah Institute of Physics, The Hebrew University, Jerusalem 91904, Israel }


\begin{abstract}
The out-of-equilibrium excess conductance of electron-glasses $\Delta$G(t)
typically relaxes with a logarithmic time-dependence. Here it is shown that
the log(t) relaxation of a weakly-disordered In$_{\text{x}}$O films
crosses-over asymptotically to an exponential dependence $\Delta$G(t)$\propto
$exp\{-[t/$\tau$($\infty$)]\}. This allows assigning a well-defined
relaxation-time $\tau$($\infty$) for a given system-disorder (characterized by
the Ioffe-Regel parameter k$_{\text{F}}\ell$). Near the metal-insulator
transition, $\tau$($\infty$) obeys the scaling relation $\tau$($\infty
$)$\propto$[(k$_{\text{F}}\ell$)$_{\text{C}}-$ k$_{\text{F}}\ell$] with the
same critical disorder (k$_{\text{F}}\ell$)$_{\text{C}}$ where the
zero-temperature conductivity of this system vanishes. The latter defines the
position of the disorder-driven metal-to-insulator transition (MIT) which is a
quantum-phase-transition. In this regard the electron-glass differs from
classical-glasses such as the structural-glass and spin-glass. The ability to
experimentally assign an unambiguous relaxation-time allows us to demonstrate
the steep dependence of the electron-glass dynamics on carrier-concentration.

\end{abstract}
\maketitle

\section{Introduction}

Slow relaxation is a widespread phenomenon manifested in a variety of physical
systems. A temporal law exhibited in many such instances is the
stretched-exponential (SE) where a measurable M relaxes with time t as:
M$\propto$exp[-(t/$\tau$)$^{\beta}$] with 0%
$<$%
$\beta$%
$<$%
1 and $\tau$ a typical relaxation time. Several approaches were offered to
account for the physical origin of this time-dependence \cite{1,2,3,4,5,6}.
Phenomenologically however, the SE may be ascribed to a distribution of
relaxation-times P($\tau$) that in turn, determines the value of $\beta$ and
$\tau$.

Another form of relaxation that is ubiquitous in disordered systems is a
logarithmic law M$\propto$-log(t) \cite{7,8,9,10,11,12,13}. Technically, a
logarithmic time dependence may be regarded as a special case of the
stretched-exponential law (when $\beta\rightarrow$0). A log(t) relaxation
conforms to a an extended range of relaxation-times P($\tau$) while $\beta$=1
is the case for a narrow distribution \cite{4}.

Demonstrating experimentally a log(t) dependence is more demanding than a
stretched-exponential in that one has no freedom in fitting; no parameter is
involved in verifying a logarithmic form whereas three parameters are
typically used to fit a SE to measured data. A log(t) relaxation is also
special in being limited to intermediate times; it ought to cross over to a
different form for both short and long times \cite{7}.

A log(t) relaxation has been claimed in a number of diversified phenomena;
structural recovery from a heat-shock \cite{7}, auto-correlations in
spin-glass \cite{8}, magnetization-relaxation \cite{9}, relaxation of
levitation force \cite{10}, exchange kinetics in polymers \cite{11}, and
dynamics of glass-forming systems \cite{12}. In some experiments the
logarithmic relaxation was monitored for more than five decades \cite{14}, but
the expected transition to the asymptotic behavior has so far escaped
detection. The main reasons for the difficulty in reaching this limit is that
the magnitude of the relaxed part of the observable is usually small to begin
with, and most of it dissipates in the early stage of the relaxation. The
finite signal-to-noise ratio and drift of instruments make it hard to identify
the expected asymptotic behavior when the P($\tau$) distribution has
components that extend over many hours.

The experiments presented in this work demonstrate that the transition from
logarithmic to exponential relaxation may consistently be observed in
electron-glasses with short relaxation-times. These observations were made on
Anderson-insulating amorphous indium-oxide films In$_{\text{x}}$O with the
lowest carrier-concentration \textit{N} yet reported to exhibit intrinsic
electron-glass features. Data for the exponential-relaxation regime allow us
to uniquely determine the relaxation-time $\tau$($\infty$) that presumably
represents the high-end of P($\tau$). Using samples with different degrees of
disorder it is shown that $\tau$($\infty$) appears to vanish at the
metal-insulator critical-point while obeying a scaling relation characteristic
of a quantum-phase-transition.

\section{Experimental}

\subsection{Sample preparation and characterization}

Samples used in this study were thin 200~\AA \ thick films of In$_{\text{x}}%
$O. These were made by e-gun evaporation of 99.999\% pure In$_{\text{2}}%
$O$_{\text{3}}$ onto room-temperature Si wafers in a partial pressure of
3x10$^{\text{-4}}$mBar of O$_{\text{2}}$ and a rate of 0.3$\pm$0.1\AA /s. The
Si wafers (boron-doped with bulk resistivity $\rho\leq$2x10$^{\text{-3}}%
\Omega$cm) were employed as the gate-electrode in\ the field-effect and
gate-excitation experiments. The samples were deposited on a SiO$_{\text{2}}$
layer (2$\mu$m thick) that was thermally-grown on these wafers and acted as
the spacer between the sample and the conducting Si:B substrate.

The as-deposited films had sheet-resistance R$_{\square}$%
$>$%
500M$\Omega$ at room-temperature. They were then thermally treated. This was
done by stages; the samples were held at a constant temperature T$_{\text{a}}$
starting from T$_{\text{a}}\approx$320 for 20-30 hours then T$_{\text{a}}$ was
raised by 5-10K at the next stage. This was repeated until the desired
R$_{\square}$ was attained (see \cite{17} for fuller details of the
thermal-annealing process). This process yielded samples with R$_{\square}%
$=20-40k$\Omega$ that at T$\approx$4K spanned the range of 100k$\Omega$ to
90M$\Omega$. A maximum T$_{\text{a}}\approx$355K and 14 days of treatment were
required to get the lowest resistance used in the study. The
carrier-concentration \textit{N} of these samples, measured by the Hall-Effect
at room-temperatures, was in the range \textit{N}$\approx$(8.5$\pm
$0.3)x10$^{\text{18}}$cm$^{\text{-3}}$.

The motivation for choosing a low-\textit{N} version of In$_{\text{x}}$O for
these experiments was the observation that the electron-glass dynamics becomes
faster as their carrier-concentration falls below \textit{N}$\lesssim
$4x10$^{\text{19}}$cm$^{\text{-3}}$ \cite{15,16} while, all other things being
equal, their excess conductance $\Delta$G is more conspicuous \cite{17}. These
expectations were borne out in our experiments which made it possible to
quantify the system dynamics as it approaches the quantum phase transition.

\subsection{Measurement techniques}

Conductivity of the samples was measured using a two-terminal ac technique
employing a 1211-ITHACO current preamplifier and a PAR-124A lock-in amplifier.
Measurements were performed with the samples immersed in liquid helium at
T$\approx$4.1K held by a 100 liters storage-dewar. This allowed up to two
months measurements on a given sample while keeping it cold. These conditions
are essential for measurements where extended times of relaxation processes
are required at a constant temperature, especially when running multiple
excitation-relaxation experiment on a given sample.

The gate-sample voltage (referred to as V$_{\text{g}}$ in this work) in the
field-effect measurements was controlled by the potential difference across a
10$\mu$F capacitor charged with a constant current. The rate of change of
V$_{\text{g}}$ is determined by the value of this current. The range of
V$_{\text{g}}$ used in this study reached in some cases $\pm$60V which is
equivalent to the $\pm$15V used in previous studies \cite{13} where the
gate-sample separation was 0.5$\mu$m as compared with the 2$\mu$m
SiO$_{\text{2}}$ spacer used here.

The ac voltage bias in conductivity measurements was small enough to ensure
near-ohmic conditions. The voltage used in the relaxation experiments was
checked to be in the linear response regime by plotting the current-voltage
characteristics of each sample. Figure 1 illustrates the dependence of the
conductance on the applied field for typical samples.%
\begin{figure}[ptb]%
\centering
\includegraphics[
height=2.316in,
width=3.039in
]%
{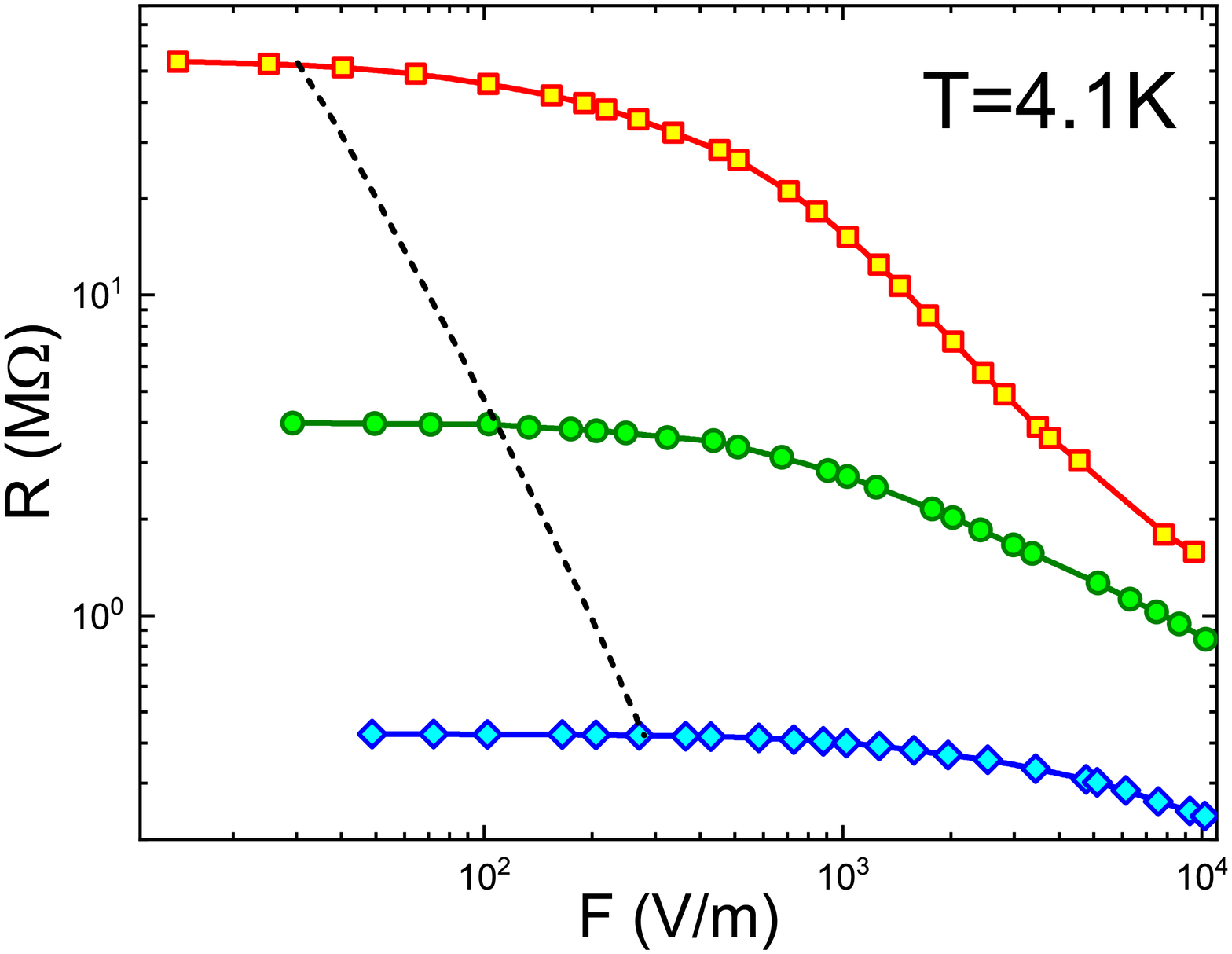}%
\caption{The dependence of sample resistance on the applied field for three
typical In$_{\text{x}}$O films, all with identical dimensions of 1x1mm. The
dashed line connects points on each R(F) curve that their deviation from Ohmic
behavior exceeds the experimental error.}%
\end{figure}

\subsection{Results and discussion}

Figure 2 shows the dependence of the conductance G on gate-voltage
V$_{\text{g}}$ for two of the studied In$_{\text{x}}$O samples. Each
G(V$_{\text{g}}$) plot show two main features; an asymmetric component
characterized by $\partial$G(V$_{\text{g}}$)/$\partial$V$_{\text{g}}$%
$>$%
0, in the entire range of -50V to +50V, that reflects the increased
thermodynamic density-of-states with energy (the thermodynamic field-effect),
and a cusp-like dip centered at V$_{\text{g}}$=0 where the system was allowed
to relax before sweeping the gate voltage (the memory-dip). Note that $\Delta
$G/G$_{\text{0}}$ for the 45M$\Omega$ sample is $\simeq$30\% - the highest
value ever reported for any electron-glass with comparable R$_{\square}$ at
T$\approx$4K and measured with a nominal sweep-rates in the range of
(0.3-1)V/s. Similarly, the 105k$\Omega$ sample has $\Delta$G/G$_{\text{0}%
}\simeq$0.3\%, still large enough for the memory-dip to stand out despite the
sloping thermodynamic field--effect. Note that the width $\Gamma$ of the
memory-dip of these samples is quite narrow consistent with their low
\textit{N }\cite{15}. Samples with different degrees of disorder, using this
batch of films, were used to monitor how, following identical excitation and
measurement conditions, the system approaches its equilibrium state.%
\begin{figure}[ptb]%
\centering
\includegraphics[
height=2.1248in,
width=3.039in
]%
{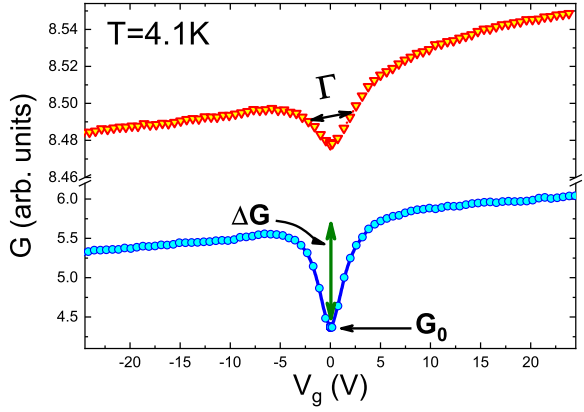}%
\caption{Conductance versus gate voltage for two of the In$_{\text{x}}$O
samples used in the study. Data for G(V$_{\text{g}}$) were taken with a sweep
rate of 0.5/s for both samples. The typical width of the memory dip $\Gamma$
and the definition of $\Delta$G and G$_{\text{0}}$ are marked with arrows. The
memory-dip relative magnitudes $\Delta$G/G$_{\text{0}}$ are $\approx$0.3\% and
$\approx$30\% for the samples with R$_{\square}$=105k$\Omega$ and 45M$\Omega$
respectively.}%
\end{figure}

An effective and reproducible way to excite the system is the `gate-protocol'
in which the excitation is affected by switching the gate-voltage
V$_{\text{g}}$ from its equilibrium value V$_{\text{eq}}$ to a new one
V$_{\text{n}}$. This takes the system out of equilibrium and this is reflected
in the appearance of a time-dependent excess-conductance $\Delta$G(t). The
minimum time the sample was equilibrated under V$_{\text{eq}}$ in this study
was 12 hours. In all the experiments described here the total gate-voltage
swing
$\vert$%
V$_{\text{n}}$-V$_{\text{eq}}$%
$\vert$
was much larger than the memory-dip width $\Gamma$ defined in Fig.2. The
relaxation to the equilibrium under the newly established V$_{\text{n}}$ was
monitored through the measured $\Delta$G(t). In previously studied
electron-glasses, this invariably led to $\Delta$G(t)$\propto$-log(t)
persisting over three or more decades. The same is true for the relaxation law
observed in other experiments where logarithmic relaxation was observed
\cite{7,9,10,12}. In the electron-glasses used here however, the log(t)
dependence was only observed for the first two decades of the relaxation. A
clear deviation from a logarithmic law to another time dependence was
consistently observed in all samples of this and three other prepared batches
of In$_{\text{x}}$O. Figure 3 shows the results of the gate-protocol for an
insulating sample with R$_{\square}$=1.5M$\Omega$.%
\begin{figure}[ptb]%
\centering
\includegraphics[
height=2.1404in,
width=3.039in
]%
{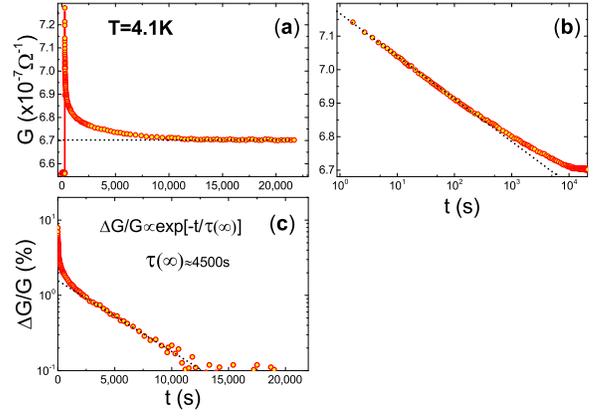}%
\caption{Results of using the gate-protocol (see text) on a sample with
R$_{\square}$=1.5M$\Omega.$ (\textbf{a}) Conductance as function of time;
after $\approx$50 seconds of monitoring G under V$_{\text{g}}$=0V the
gate-voltage was swept to V$_{\text{g}}$=20V at a rate of 10V/s. The dashed
line is G($\infty$), the asymptotic value of the conductance. G($\infty$)
differs from G(V$_{\text{eq}}$) due to the component of the thermodynamic
field-effect. (\textbf{b}) Conductance relaxation starting from the time
V$_{\text{g}}$=20V was established showing the extent of the lag(t) dependence
(delineated by the dashed line). (\textbf{c}) The plot of G(t)-G($\infty$)
demonstrating an exponential relaxation behavior (dashed line is best fit).}%
\end{figure}
As the figure illustrates, the excess-conductance initially follows $\Delta
$G(t)$\propto$-log(t) for a time period lasting up to $\approx$250s, (Fig.3b),
then $\Delta$G deviates from the logarithmic law, showing a tendency for the
conductance to saturate. It turns out that for t$\gtrsim$1500s the excess
conductance $\Delta$G(t) for this sample may be well fitted to a
stretched-exponential dependence with $\beta\approx$0.95$\pm$0.1 (Fig.3c).
Within the experimental error this $\beta$ is close enough to unity.
Accordingly, we fit the data in this regime assuming $\beta\equiv$1, that is:
$\Delta$G(t)$\propto$exp$[$-t/$\tau$]. It seems that for these samples, the
low-end of the rate-distribution is narrow enough to make it impossible to
distinguish $\beta$ from unity.

Qualitatively similar results were obtained for the gate-protocol applied to a
105k$\Omega$ sample, the sample with the lowest degree of disorder that was
used for this purpose. These results are shown in Fig.4:%
\begin{figure}[ptb]%
\centering
\includegraphics[
height=2.1248in,
width=3.039in
]%
{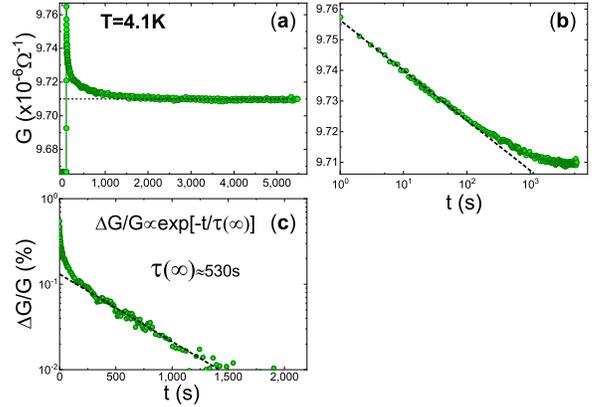}%
\caption{Same as in Fig.3 but for a sample with R$_{\square}$=105k$\Omega$.
Note the difference in the time-scale as compared with Fig.2 and the constant
G(t) for t$\eqslantgtr$2000s.}%
\end{figure}
The figure shows again the limited range over which the excess conductance
follows a log(t) dependence and the transition to an exponential-relaxation
regime at later times. Apart from the lower signal-noise ratio (due to the
smaller $\Delta$G/G$_{\text{0}}$ associated with the lower R$_{\square}$
sample), the main difference between the data in Fig.4 and Fig.3 is in the
relaxation-time $\tau$($\infty$). This time turns out to be independent of the
particular value of V$_{\text{eq}}$ and V$_{\text{n}}$, as long as
$\vert$%
V$_{\text{n}}$-V$_{\text{eq}}$%
$\vert$
is larger than the width of the memory dip$~\Gamma$. This aspect was tested in
the sample of Fig.4 using eight different runs of the gate-protocol with
20$\eqslantless$%
$\vert$%
V$_{\text{n}}$-V$_{\text{eq}}$%
$\vert$%
$\eqslantless$ 90 volts, and in seven runs for a sample with R$_{\square}%
$=120k$\Omega$ using voltage-swings in the range of 40-80 volts. In either
case, the same behavior of G(t) and the same value of $\tau$($\infty$) was
obtained. Note that it is the relatively short equilibration time of these
samples that makes these repeated measurements on the given sample practical
(in addition to the benefit of short relaxation times in minimizing the
problem of instruments drift).

In the range of equilibration times used in this work, the relaxation time
$\tau$($\infty$) is also independent of the time the sample was held under
V$_{\text{eq}}~$before switching to V$_{\text{n}}$. Figure 5 compares the
results of the gate-protocol for two different times the system was allowed to
equilibrate under V$_{\text{eq}}$.%
\begin{figure}[ptb]%
\centering
\includegraphics[
height=2.1309in,
width=3.039in
]%
{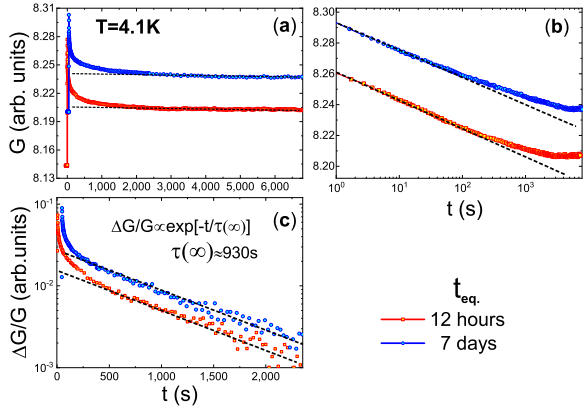}%
\caption{{}Comparing the results of the gate-protocol (with V$_{\text{eq}}$=0V
and V$_{\text{n}}$=46V) for a sample with R$_{\square}$=120k$\Omega$ using two
different equilibration times, t$_{\text{{\protect\small eq}}}$. (a) G(t) data
for the entire duration of the protocol with V$_{\text{eq}}$=0V, V$_{\text{n}%
}$=46V, and sweep rate of 10V/s. The curves are displaced for clarity. (b)
G(t) for the relaxation. The origin of time is taken as the instance when
V$_{\text{g}}$~reached V$_{\text{n}}$. Note the similar appearance of the
curves (dashed lines are the respective best-fits). }%
\end{figure}

On the other hand, the relaxation time $\tau$($\infty$) does depend on
R$_{\square}$ as it may already been noticed by comparing the time-scales in
Fig.3 and Fig.4. In fact, the relaxation time of these electron-glasses
decreases systematically with their resistance and, as will be shown below,
eventually vanishes when the sample resistance is sufficiently small.

This trend has implication on the origin of the slow relaxation in these
systems. The reduction of $\tau$($\infty$) with R$_{\square}$ is achieved in
In$_{\text{x}}$O by thermal-annealing. The changes in the structural
properties of the material during the annealing process were extensively
studied in \cite{18} by electron-diffraction, energy-dispersive spectroscopy,
x-ray interferometry, and optical techniques. These studies revealed that the
change in the resistance from the as-deposited deeply insulating state all the
way to the metallic regime is mainly due to increase of the material
\textit{density}. In particular, the samples retained their amorphous
structure and composition throughout the entire process \cite{18}. Moreover,
the dynamics associated with structural changes monitored during annealing and
recovery of the samples was qualitatively different than that of the
electron-glass and did not change its character throughout the entire range of
disorder \cite{16}. The diminishment of $\tau$($\infty$) with R$_{\square}$
cannot then be identified with the elimination of some peculiar structural
defects that were hypothesized as being responsible for the slow relaxation
and other associated nonequilibrium transport phenomena in electron-glasses
\cite{19}.

On the other hand, as all the samples studied here had essentially the same
\textit{N}, the difference in their relaxation times with their different
resistance may hint on the role of \textit{quenched} disorder. This motivated
us to measure $\tau$($\infty$) in films of the same preparation batch that
underwent different degrees of thermal annealing to fine-tune their disorder.

A dimensionless measure of disorder that is often used in metal-insulator and
superconducting-insulator studies is the Ioffe-Regel parameter k$_{\text{F}%
}\ell$. This may be estimated by k$_{\text{F}}\ell$=(3$\pi^{\text{2}}%
$)$^{\text{2/3}}\text{\textperiodcentered}\hbar$\textperiodcentered
$\sigma_{\text{RT}}\text{\textperiodcentered}e^{\text{-2}}$\textperiodcentered
\textit{N}$^{\text{-1/3}}$ where $\sigma_{\text{RT}}$ is the sample
conductivity at room-temperature. We use this measure to plot in Fig.6 the
dependence of $\tau$($\infty$) on disorder. The figure shows that $\tau
$($\infty$) appears to vanish at (or near) a specific (k$_{\text{F}}\ell
$)$_{\text{C}}$. Interestingly, this point in disorder coincides with the
metal-insulator transition (MIT) for In$_{\text{x}}$O determined \cite{20} in
the same version of In$_{\text{x}}$O used in the current study (in terms of
having a similar carrier-concentration \textit{N}).%
\begin{figure}[ptb]%
\centering
\includegraphics[
height=2.2917in,
width=3.039in
]%
{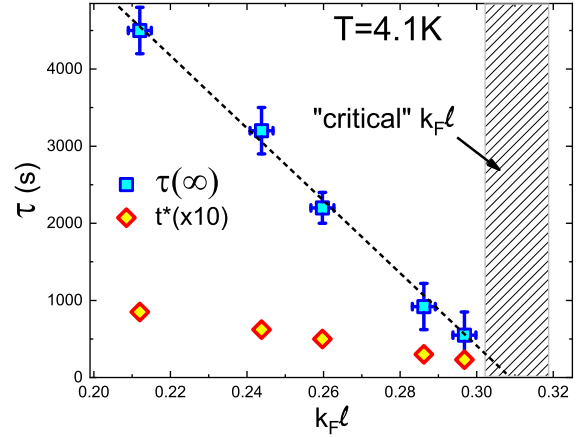}%
\caption{The dependence of the relaxation times $\tau$($\infty$) (defined in
Figs.3,4,5) and t* (see definition in the text) on the disorder parameter
k$_{\text{F}}\ell$ near the critical-regime of the metal-to-insulator
transition (marked by the hatched area).}%
\end{figure}

The diminishment of the relaxation-time as the metallic phase is approached
from the insulating phase is in line with the notion that glassy behavior is
an inherent feature of Anderson-insulators
\cite{21,22,23,24,25,26,27,28,29,30,31}.

The dynamics at the metal-insulator critical-point (defined by disorder where
the dc conductivity vanishes as T$\rightarrow$0), is likely slower than deeper
into the metallic phase; sections of the sample may still be insulating and
their slow dynamics could influence the transport in the just-percolating
diffusive channel \cite{32,33}. Glassy effects may then persist some distance
into the diffusive phase but their magnitude will be small and the observed
dynamics will speed-up sharply as k$_{\text{F}}\ell$ increases past the
critical disorder.

The trend for faster dynamics as k$_{\text{F}}\ell\rightarrow$(k$_{\text{F}%
}\ell$)$_{\text{C}}$ is also reflected during the log(t) relaxation regime.
This may be quantified by choosing (arbitrarily) a time-scale t* defined
through 2G(t*)=G(1)-G($\infty$). This yielded t* of 23-85s for the samples
with k$_{\text{F}}\ell$ in the range 0.29-0.22 respectively. The dependence of
t* on disorder is shown in Fig.6.

The scaling relation $\tau$($\infty$)$\propto$%
$\vert$%
k$_{\text{F}}\ell$-(k$_{\text{F}}\ell$)$_{\text{C}}$%
$\vert$%
$^{\nu}$ with $\nu\approx$1 obeyed by the relaxation-time (Fig.6) is
essentially the same as the scaling behavior of the zero-temperature
conductance $\sigma$(0)$\propto$%
$\vert$%
k$_{\text{F}}\ell$-(k$_{\text{F}}\ell$)$_{\text{C}}$%
$\vert$%
$^{\nu}$ reported in \cite{20} for the MIT of this material. The transition to
the electron-glass phase thus emerges as a \textit{quantum-phase transition}
rather than following the scenario conceived by Davies et al \cite{21}. These
authors anticipated a glass temperature by analogy with the spin-glass problem
\cite{21}. Instead, the experiments suggest that the glassy effects are just
finite-temperature manifestations of a zero-temperature disorder-driven
transition in the same vein that an exponential temperature dependence of the
conductance attests to the Anderson-insulating phase.

Our results emphasize the dominant role of the disorder in controlling the
slow relaxation of electron-glasses. Note that the present system has the
lowest carrier-concentration among the electron-glasses studied to date that,
in turn, it is also the \textit{least disordered} electron-glass; the disorder
required to make the system insulating is smaller the smaller is the
Fermi-energy (or equivalently \textit{N}). It is due to these conditions that
it was possible to observe the transition to the ergodic regime of relaxation.

As the transition is approached from the insulating side, the relaxation time
concomitantly decreases with the reduced quenched-disorder thus curtailing the
time-scale over which electron-glass attributes may be observed by
conventional transport techniques.

It is illuminating to compare the glass dynamics reported here using
In$_{\text{x}}$O with carrier-concentration \textit{N}$\approx$(8.7$\pm
$0.2)x10$^{\text{18}}$cm$^{\text{-3}}$ with the dynamics observed in
measurements made on another version of the same material with \textit{N}%
$\approx$(7.5$\pm$0.5)x10$^{\text{19}}$cm$^{\text{-3}}$ \cite{34}. In the
present work the log(t) relaxation was shown to persist for two decades in
time whereas in the In$_{\text{x}}$O version with \textit{N}$\approx$(7.5$\pm
$0.5)x10$^{\text{19}}$cm$^{\text{-3}}$ the logarithmic law was still
observable for \textit{two more} decades in time \cite{34}. Figure 7 shows the
magnitude of the memory-dip versus the resistivity of these versions of
In$_{\text{x}}$O that structurally are essentially identical except that due
to a small change of composition they differ in carrier-concentration. The
figure demonstrates that, all other things being equal, the relative magnitude
of the out of equilibrium excess-conductance is considerably larger in the
currently studied version of In$_{\text{x}}$O which, in turn, means that its
much faster dynamics is not due to the limiting small signal and conversely,
the higher-N version slower dynamics may actually be more than two decades. In
other words, going down one decade in carrier-concentration the electron-glass
dynamics becomes faster by \textit{at least} two decades.%
\begin{figure}[ptb]%
\centering
\includegraphics[
height=2.3514in,
width=3.039in
]%
{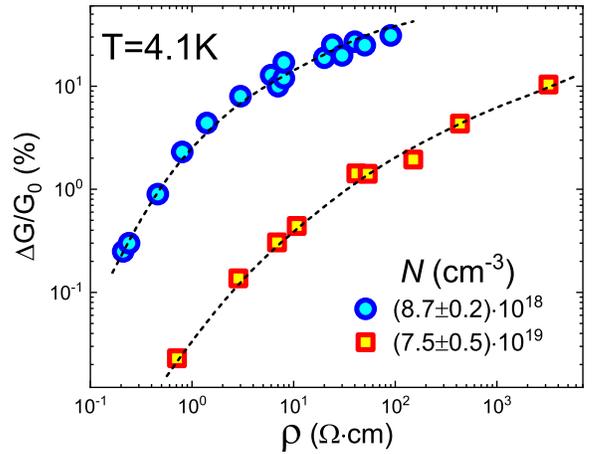}%
\caption{The relative magnitude of the memory-dip (see Fig.2 for definition)
as function of resistivity for two versions of In$_{\text{x}}$O films measured
under the same conditions. Dashed lines are guides for the eye.}%
\end{figure}

This result may help to put in perspective the failure of some
Anderson-insulators, most notably, Si and GaAs, to exhibit intrinsic
electron-glass features. The absence of lightly-doped semiconductors from the
list of systems that show glassy effects made it hard for some researchers to
accept that the electron-glass is just another property of the interacting
Anderson-insulating phase. It has been suggested \cite{18} that the failure to
observe a memory-dip in Anderson-insulators with low carrier-concentration is
their fast relaxation-times; to resolve the memory-dip by a field-effect
measurement requires relaxation-time larger than the sweep-time used in the
G(V$_{\text{g}}$) measurements. It was then conjectured that the
relaxation-time of Anderson insulators sharply depends on \textit{N} so only
systems with large carrier-concentrations exhibit a memory-dip \cite{18}. The
finite relaxation time of the low-\textit{N} version of In$_{\text{x}}$O
studied here gives an unambiguous experimental point of reference in support
of this conjecture. We expect that if a version of In$_{\text{x}}$O with
\textit{N}$\approx$10$^{\text{17}}$cm$^{\text{-3}}$ (typical for lightly-doped
semiconductors in their hopping regime) could be made, and if its resistance
at low temperatures could be made to be compatible with field-effect
measurements, then its relaxation time will be too short to allow a memory-dip
to be observed just as is the case for Si and GaAs.

Carrier-concentration is evidently an important parameter in determining the
relaxation time of electron-glasses but it should be emphasized that other
factors may play a significant role. In particular, the electron-phonon
coupling is obviously an essential ingredient in energy relaxation processes
of the electronic system. Further reduction of transition-rates relative to
the "bare" rates controlled by disorder may occur for non-local interactions.
These may bring into play additional constraints \cite{3} as well as effects
related to coupling of the tunneling charge to other degrees of freedom
(polaronic-effects \cite{33}, and the orthogonality-catastrophe
\cite{35,36,37}).

It is yet unclear how important is the long-range Coulomb interaction to the
observed slow dynamics. Relaxation times that extend over thousands of seconds
are observable at temperatures where the hopping-length, which is the
effective screening-length in the insulating regime, is of the order of
$\approx$10nm. It seems therefore that, in addition to strong enough quenched
disorder, medium-range interactions may be sufficient to account for the
relaxation times observed in the experiments. This issue is now under
investigation by using a metallic ground-plane in proximity to the sample to
modify the long-range Coulomb interaction in a controlled way.

Illuminating discussions with A. Zaccone and with A. Vaknin are gratefully
acknowledged. This research has been supported by a grant No 1030/16
administered by the Israel Academy for Sciences and Humanities.


\begin{thebibliography}{99}                                                                                               %


\bibitem {1}J. C. Phillips, Rep. Prog. Phys. \textbf{59}, 1133 (1996).

\bibitem {2}E. W. Montroll, J. T. Bendler, J. Stat. Phys. \textbf{34},129 (1983).

\bibitem {3}R. G. Palmer, D. L. Stein, E. Abrahams, and P. W. Anderson, Phys.
Rev. Lett., \textbf{53}, 958 (1984).

\bibitem {4}D. C. Johnston, hys. Rev. B \textbf{74}, 184430 (2006).

\bibitem {5}J. S. Langer, S. Mukhopadhyay, Phys. Rev. E \textbf{77}, 061505 (2008).

\bibitem {6}P. Grassberger, I. Procaccia, J. Chem. Phys. \textbf{77}, 6281 (1982).

\bibitem {7}G. Mih\`{a}ly and L. Mih\`{a}ly, Phys. Rev. Lett., \textbf{52},
149 (1984).

\bibitem {8}Daniel S. Fisher and David A. Huse, Phys. Rev. B \textbf{38}, 386 (1988).

\bibitem {9}Javier Tejada, and Xixiang Zhang, Journal of Magnetism and
Magnetic Materials, \textbf{140-144} 1815 (1995).

\bibitem {10}A. B. Riise, T. H. Johansen, H. Bratsberg, and Z. J. Yang, Appl.
Phys. Lett. \textbf{60}, 2294 (1992).

\bibitem {11}W. G\"{o}tze and M. Sperl, Phys. Rev. E \textbf{66}, 011405 (2002).

\bibitem {12}Reidar Lund, Lutz Willner, and Dieter Richter, Macromolecules,
\textbf{39}, 4566 (2006).

\bibitem {13}A. Vaknin, Z. Ovadyahu, and M. Pollak, Phys. Rev. B \textbf{61},
6692 (2000).

\bibitem {14}Z. Ovadyahu, Phys. Rev. B \textbf{73}, 214204 (2006).

\bibitem {15}A. Vaknin, Z. Ovadyahu, and M. Pollak, Phys. Rev. Lett.,
\textbf{81}, 669 (1998).

\bibitem {16}Z. Ovadyahu, Phys. Rev. B \textbf{95}, 214207 (2017).

\bibitem {17}Z. Ovadyahu, Phys. Rev. B \textbf{97}, 054202 (2018).

\bibitem {18}Z. Ovadyahu, Phys. Rev. B. 95, 134203 (2017).

\bibitem {19}A. L. Burin, V. I. Kozub, Y. M. Galperin, and V. Vinokur, J.
Phys. C \textbf{20}, 244135 (2008) and references therein.

\bibitem {20}D. Shahar and Z. Ovadyahu, Phys. Rev. B \textbf{46}, 10917
(1992); U. Givan and Z. Ovadyahu, Phys. Rev. B \textbf{86}, 165101 (2012).

\bibitem {21}J. H. Davies, P. A. Lee, and T. M. Rice, Phys. Rev. Letters,
\textbf{49}, 758 (1982); M. Gr\"{u}newald, B. Pohlman, L. Schweitzer, and D.
W\"{u}rtz, J. Phys. C, \textbf{15}, L1153 (1982); J. H. Davies, P. A. Lee, and
T. M. Rice, Phys. Rev. B \textbf{29}, 4260 (1984).

\bibitem {22}M. Pollak and M. Ortu\~{n}o, Sol. Energy Mater., \textbf{8}, 81
(1982); M. Pollak, Phil. Mag. B\textbf{\ 50}, 265 (1984).

\bibitem {23}G. Vignale, Phys. Rev. B\textbf{\ 36}, 8192 (1987).

\bibitem {24}C. C. Yu, Phys. Rev. Lett., \textbf{82}, 4074 (1999).

\bibitem {25}M. M\"{u}ller and L. B. Ioffe, Phys. Rev. Lett. \textbf{93},
256403 (2004).

\bibitem {26}Vikas Malik and Deepak Kumar, Phys. Rev. B \textbf{69}, 153103 (2004).

\bibitem {27}R. Grempel, Europhys. Lett., \textbf{66,} 854 (2004).

\bibitem {28}Eran Lebanon, and Markus M\"{u}ller, Phys. Rev. B\textbf{\ 72},
174202 (2005); M. M\"{u}ller and E. Lebanon, J. Phys. IV France, \textbf{131},
167 (2005).

\bibitem {29}Ariel Amir, Yuval Oreg, and Yoseph Imry, Phys. Rev. B
\textbf{77}, 165207 (2008); Ariel Amir, Yuval Oreg, and Yoseph Imry, Annu.
Rev. Condens. Matter Phys. \textbf{2,} 235 (2011); Y. Meroz, Y. Oreg and Y.
Imry, EPL, \textbf{105}, 37010 (2014).

\bibitem {30}M. Pollak, M. Ortu\~{n}o and A. Frydman, "\textit{The Electron
Glass}\textbf{",} Cambridge University Press, England (2013).

\bibitem {31}Y. Meroz, Y. Oreg and Y. Imry, EPL, \textbf{105}, 37010 (2014).

\bibitem {32}Kendall Mallory, Phys. Rev. B \textbf{47}, 7819 (1993).

\bibitem {33}V. I. Kozub, Y. M. Galperin, V. Vinokur, and A. L. Burin, Phys.
Rev. B \textbf{78}, 132201 (2008).

\bibitem {34}Z. Ovadyahu, Phys. Rev. B. \textbf{90}, 054204 (2014).

\bibitem {35}Vedika Khemani, Rahul Nandkishore, and S. L. Sondhi, Nature
Physics, \textbf{11}, 560 (2015).

\bibitem {36}P. W. Anderson, Phys. Rev. Lett. \textbf{18}, 1049 (1967); A. J.
Leggett et al., Rev. Mod. Phys. \textbf{59}, 1 (1987); Z. Ovadyahu, Phys. Rev.
Lett., \textbf{99}, 226603 (2007).

\bibitem {37}D. L. Deng, J. H. Pixley, X Li, S. D. Sarma, Phys. Rev. B.
\textbf{92}, 220201(R) (2015).
\end{thebibliography}
\end{document}